\documentclass[12pt]{article}

\catcode`\@=11

\global\arraycolsep=2pt
\usepackage{amsbsy,amssymb,latexsym,amsfonts,amsmath}
\usepackage{graphicx,color}

\allowdisplaybreaks

\begin{document}
\begin{flushright}
\parbox{4.2cm}
{RUP-20-29}
\end{flushright}

\vspace*{0.7cm}

\begin{center}
{ \Large Anomalous hydrodynamics with dyonic charge}
\vspace*{1.5cm}\\
{Yu Nakayama}
\end{center}
\vspace*{1.0cm}
\begin{center}

Department of Physics, Rikkyo University, Toshima, Tokyo 171-8501, Japan

\vspace{3.8cm}
\end{center}

\begin{abstract}
We study anomalous hydrodynamics with a dyonic charge. We show that the local second law of thermodynamics constrains the structure of the anomaly in addition to the structure of the  hydrodynamic constitutive equations. In particular, we show that not only the usual $E\cdot B$ term but also $E^2 -B^2$ term should be present in the anomaly  with a specific coefficient for the local entropy production to be positive definite.
\end{abstract}

\thispagestyle{empty} 

\setcounter{page}{0}

\newpage

\section{Introduction}
The existence of anomaly in quantum field theories had been well-known, but it was only recently that its consequence in the hydrodynamic regime was revealed \cite{Son:2009tf} (see also \cite{Sadofyev:2010pr}\cite{Neiman:2010zi}\cite{Loganayagam:2012pz}\cite{Banerjee:2012iz}\cite{Jensen:2012jy}\cite{Bhattacharya:2012zx}\cite{Haehl:2013hoa}\cite{Golkar:2015oxw}\cite{Glorioso:2017lcn}). It predicts universal new kinetic coefficients in the hydrodynamic constitutive equations for the  current with the triangle anomaly:
\begin{align}
\partial^\mu J_\mu = c F^{\mu\nu} \tilde{F}_{\mu\nu} \ . 
\end{align}
The key idea was that under the presence of the triangle anomaly, it is imperative to introduce a new kinetic coefficient proportional to the vorticity in order for the local second law of thermodynamics (i.e. the positivity of the local entropy production) to hold under the presence of the external gauge field that causes the anomaly. The new kinetic coefficients cause no dissipation and they are fixed by the triangle anomaly of the microscopic theory.

In this paper, we would like to ask the question about what will happen if we try to gauge the current (with anomaly) not in the electric way but in the more generic dyonic way. By this, we mean that the current is coupled with the external gauge field both electrically and magnetically. In particular, we would like to study the consequence in the hydrodynamic regime.

We show that the local second law of thermodynamics constrains the structure of the anomaly in addition to the structure of the  hydrodynamic constitutive equation. In particular,  we will see that in order to maintain the local second law of thermodynamics, the anomaly must include the extra term
\begin{align}
\partial^\mu J_\mu = c F^{\mu\nu} \tilde{F}_{\mu\nu} + \tilde{c} F^{\mu\nu}F_{\mu\nu} \ , 
\end{align}
whose coefficient is fixed by the magnetic gauging.

In conventional quantum field theories, we rarely see $F^{\mu\nu}F_{\mu\nu}$ term in the anomalous conservation law although it is consistent \cite{Nakayama:2018dig} (in the Wess-Zumino sense). Our results, however, suggest that its appearance is crucial to assure the positivity of the local entropy production in the hydrodynamic regime if we try to gauge the current in the dyonic way. 

\section{Anomalous hydrodynamics with dyonic charge}
Let us consider an anomalous current $J_\mu$ in a $1+3$ dimensional relativistic quantum field theory and try to couple it with the external gauge field in a dyonic way. It means that under the presence of the external gauge field (with the field strength denoted by $F_{\mu\nu}$ and its dual $\tilde{F}_{\mu\nu} = \frac{1}{2}\epsilon_{\mu\nu\rho\sigma}F^{\rho\sigma}$), the conservation equations for the energy-momentum tensor $T^{\mu\nu}$ and the anomalous current $J^\mu$ (in the covariant form) become
\begin{align}
\partial_\mu T^{\mu\nu} &= (F^{\nu\rho} + \beta \tilde{F}^{\nu\rho}) J_\rho \cr
\partial_\mu J^\mu &= c E^\mu B_\mu + \tilde{c}(E^2-B^2) \ .  \label{anoc}
\end{align}
Here, we have introduced the electric and magnetic fields defined with respect to the rest frame of the fluid from the velocity vector field $u^\mu$ (with the normalization given by  $u_\mu u^\mu=-1$): $E^\mu = F^{\mu\nu} u_\nu$, $B^\mu = \frac{1}{2}\epsilon^{\mu\nu\rho\sigma} u_\nu F_{\rho\sigma}$. Usual chiral anomaly does not introduce the second term $\tilde{c}(E^2-B^2) = -\frac{\tilde{c}}{2}F^{\mu\nu} F_{\mu\nu}$, but we will see that it plays a crucial role to maintain the local second law of thermodynamics. The parameter $\beta$ is an arbitrary real number for our classical treatment.  

We also assume that the external field strength satisfies the``Bianchi identity"
\begin{align}
\partial^\mu (\beta F_{\mu\nu} - \tilde{F}_{\mu\nu}) = 0 \ , \label{Bianchi}
\end{align}
which means that our charge lattice is one-dimensional. A difficulty without imposing the ``Bianchi identity" will be discussed later. Note that imposing this condition is consistent with the (anomalous) conservation laws \eqref{anoc}. Note also that even in the $\beta =0$ case studied in the literature \cite{Son:2009tf}\cite{Neiman:2010zi}, we could not have obtained the local second law of thermodynamics if we did not  impose the Bianchi identity (i.e. \eqref{Bianchi} with $\beta=0$).\footnote{As a further technical remark, in may cases, we also have to use the ``Bianchi identity" to transform the anomalous conservation equations into a covariant form as  \eqref{anoc}.}

In hydrodynamics, we expand the energy-momentum tensor and the current with respect to derivatives of the velocity vector $u^\mu$ and the local temperature $T$ and the (dyonic) chemical potential $\mu$. Since we have assumed that our charge lattice is one-dimensional, we have only one chemical potential here.

At the zero-th order (i.e. in the ideal fluid limit) with the use of the projector $P_\mu^\nu = \delta_{\mu}^\nu + u_\mu u^\nu$, we have 
\begin{align}
T^{(0)}_{\mu\nu} &= \epsilon u_\mu u_\nu + p P_{\mu\nu} \cr 
 J^{(0)}_{\mu} & = n u_\mu \cr
s^{(0)}_{\mu} & = s u_\mu \ ,  
\end{align}
where local energy $\epsilon$, local pressure $p$, local number density $n$ and local entropy density $s$ are functions of $\mu$ and $T$ subject to the first law of local thermodynamics $\epsilon + P = Ts + \mu n$.\footnote{They might depend on the external gauge fields $F_{\mu\nu}$. We assume $F_{\mu\nu} \sim O(p)$. Accordingly we only keep $O(p)$ in the constitutive equations for $T_{\mu\nu}$ and $J_\mu$ to discuss the first order hydrodynamics.} If we are working on conformal fluid, the (anomalous) trace formula $T^{\mu}_{\mu} = \mathcal{O}(F_{\mu\nu}^2)$ gives $\epsilon = 3 p$ up to $\mathcal{O}(F_{\mu\nu}^2)$ corrections.\footnote{This $\mathcal{O}(F_{\mu\nu}^2)$ term includes $E^\mu B_\mu$ term as well when $\beta$ is non-zero \cite{Csaki:2010rv}\cite{Nakagawa:2020gqc}\cite{Nakayama}.}

The ideal fluid equations become
\begin{align}
(u^\mu \partial_\mu) (\epsilon + p) u_\nu + (\epsilon+p) (\partial^\mu u_\mu) u_\nu + \partial_\nu p &=  n (E_\nu + \beta B_\nu) \cr
u^\mu \partial_\mu  n +  n \partial^\mu u_\mu &= 0 \ ,
\end{align}
from which (together with the ``Bianchi identity") we can derive useful relations for the vorticity vector $\omega^\mu = \frac{1}{2}\epsilon^{\mu\nu\rho\sigma} u_{\nu}\partial_\rho u_\sigma $: 
\begin{align}
\partial^\mu \omega_\mu = -\frac{2}{\epsilon + p} \omega^\mu (\partial_\mu p - n (E_\mu + \beta B_\mu)) \label{u1}
\end{align}
and the divergence of field strength:
\begin{align}
&\partial_\mu (B^\mu - \beta E^\mu) = \cr
& -2\omega^\mu (E_\mu + \beta B_\mu) + \frac{1}{\epsilon+p}(-(B^\mu-\beta E^\mu)\partial_\mu p + n (E_\mu + \beta B_\mu)(B^\mu -\beta E^\mu)) \ . \label{u2}
\end{align}

Now let us consider the first order corrections to the constitutive equations. By imposing the Landau frame condition $u^\mu T^{(1)}_{\mu\nu} = u^\mu J^{(1)}_{\mu}= 0$, we have the most generic possibility
\begin{align}
T^{(1)}_{\mu\nu} &= -(2\eta \pi_{\mu\nu} + \zeta P_{\mu\nu} (\partial^\rho u_\rho )) \cr
J^{(1)}_{\mu} & = \chi P_{\mu\nu} \partial^\nu p - T \sigma P_{\mu\nu} \partial^\nu \frac{\mu}{T} + \sigma^{(E)} (E_\mu + \beta B_\mu) +\xi \omega_\mu + \xi^{(B)} (B_\mu -\beta E_\mu)  \cr
s^{(1)}_{\mu} &= -\frac{\mu}{T} J^{(1)}_\mu + \tilde{\zeta} u_\mu (\partial \cdot u) + \tilde{\chi} P_{\mu\nu} \partial^\nu p + \tilde{\sigma}P_{\mu\nu} \partial^\nu \frac{\mu}{T}+\tilde{\sigma}^{(E)} (E_\mu + \beta B_\mu)  \cr
 &+ D \omega_\mu + D^{(B)} (B_\mu -\beta E_\mu)  \ . 
\end{align}
Here, $\pi^{\mu\nu} = P^{\mu\alpha}P^{\nu \beta}(\partial_{\alpha}  u_\beta + \partial_\beta u_\alpha - \frac{\partial^\rho u_\rho}{3}\eta_{\alpha \beta})$ is the shear tensor.

By using the anomalous conservation \eqref{anoc}, we obtain
\begin{align}
\partial_\mu s^\mu &= \frac{1}{T} \left(-T^{(1)}_{\mu\nu} D^{\mu} u^\nu + J^{(1)}_{\mu} (E^\mu + \beta B^\mu - T \partial^\mu\frac{\mu}{T}) -  c \mu E^\mu B_\mu - \tilde{c}\mu (E^2-B^2)  \right) \cr 
& + \partial^\mu (s^{(1)}_{\mu} + \frac{\mu}{T} J^{(1)}_{\mu}) \ . \label{entrod}
\end{align}
In order to realize the local second law of thermodynamics, we would like to demand that the right hand side of \eqref{entrod} is non-negative, which can be evaluated by using \eqref{u1} and \eqref{u2}. The most of the analysis follows from the earlier works \cite{Son:2009tf}\cite{Neiman:2010zi}, where we can effectively replace $E_\mu$ there with $E_\mu + \beta E_\mu$ and $B_\mu$ there with $B_\mu -\beta E_\mu$. In particular, we need to set $\tilde{\chi} = \tilde{\zeta} = \chi = \tilde{\sigma} = \tilde{\sigma}^{(E)} = 0$ with $\sigma \ge 0 $, $\eta\ge 0$, $\zeta\ge 0$. 

The most significant difference here is that the form of the anomaly consists of two terms $c E^\mu B_\mu + \tilde{c}(E^2-B^2)$, both of which do not have a positive property in the divergence of the entropy current. It means that the both contributions must be canceled from the modification of the constitutive equations. 

We actually do know that the particular combination proportional to $(E^\mu + \beta B^\mu)(B_\mu -\beta E_\mu)$ can be canceled from the studies of \cite{Son:2009tf}\cite{Neiman:2010zi} with the simple replacement we have just mentioned, where they have shown that we can set 
\begin{align}
D &= \frac{C \mu^3}{3T} \cr
D^{(B)} & = \frac{C \mu^2}{2T} \cr 
\xi & = C \left(\mu^2 - \frac{2}{3} \frac{n \mu^3}{(\epsilon + p)}\right) \cr
\xi^{(B)} & = C   \left(\mu - \frac{1}{2} \frac{n \mu^2}{(\epsilon + p)}\right) \ 
\end{align}
to cancel the term $C (E^\mu + \beta B^\mu)(B_\mu -\beta E_\mu)$.\footnote{Without affecting the cancellation, one may add further parity breaking terms (e.g. $\gamma T^2$ term in $D$) as first observed in \cite{Neiman:2010zi}.} Note that the term $\xi^{(B)}$ induces not only the chiral magnetic effect but also the ``chiral electric effect". Since these effects do not cause the entropy production, they are non-dissipative. 

We, however, see that the remaining terms in  $c E^\mu B_\mu + \tilde{c}(E^2-B^2)$ cannot be canceled or be made positive definite. This can be seen from the observation that among three basis terms of $ (B^\mu -\beta E^\mu)(B_\mu -\beta E_\mu),  (B^\mu -\beta E^\mu)(E_\mu +\beta B_\mu),  (E^\mu +\beta B^\mu)(E_\mu +\beta B_\mu)$, the bilinear $\mu (B^\mu -\beta E^\mu)(B_\mu -\beta E_\mu)$ that appears in the entropy production comes only from the anomaly and it cannot be obtained from our modification of the entropy current. It is not positive definite either because the sign of $\mu$ is unconstrained.  

Therefore the only possible way to maintain the local second law of thermodynamics seems to assume that $c$ and $\tilde{c}$ are related:
\begin{align}
\tilde{c} = -\frac{\beta}{1-\beta^2} c \ 
\end{align}
so that there is no remaining $\mu (B^\mu -\beta E^\mu)(B_\mu -\beta E_\mu)$  term by choosing $C=\frac{c}{1-\beta^2}$. Indeed, this relation of the anomaly of the dyonic current is proposed in \cite{Argyres:1995jj}\cite{Csaki:2010rv} from the $SL(2,\mathbb{Z})$ invariance, and we do know that such a theory must be consistent (by assuming the $SL(2,\mathbb{Z})$ duality of anomalous gauging).

It is worthwhile noticing that we may set the external field strength to be zero and still see the effect of the anomaly as a chiral vortex effect. The strength of the chiral vortex effect is fixed by the anomaly coefficient of the underlying theory and does not depend on the choice whether we electrically gauge it or dyonically gauge it. In our case, the artificial dependence of $\beta$ can be removed by a redefinition of $\mu$ (whose normalization we have not fixed).

\section{Discussions}
In this paper, we have studied anomalous hydrodynamics with a dyonic charge. We have seen that the local second law of thermodynamics constrains the structure of the anomaly. We have  shown that not only the usual $E\cdot B$ term but also $E^2 -B^2$ term should be present in the anomaly  with a specific coefficient. We see that this particular coefficient is precisely what we expect from the $SL(2,\mathbb{Z})$ duality as a viable possibility, but it is not immediately obvious why it would violate the local second law of thermodynamics otherwise.

The question could have been addressed with the electric gauging with $\beta=0$. If the current is anomalous with non-zero $E^2-B^2$ term, then the local second law of thermodynamics would fail. Independently of this observation, we know that the anomalous current with  $E^2-B^2$ term in conformal field theory is ``impossible" in the sense that the non-local three-point functions of conserved current that would generate the $E^2-B^2$ anomaly do not exist \cite{Nakayama:2018dig}. However, this fact does not immediately lead to the conclusion that such terms are not physical: we can for instance construct the (conformally invariant) Wess-Zumino effective action that generates the semi-local three-point functions that give rise to the $E^2-B^2$ anomaly. In this sense, it may be perplexing that the local second law of thermodynamics fails.\footnote{We often say that the usual $E\cdot B$ anomaly is invariant under time-reversal so the anomalous transport must be non-dissipative. Here, the coexistence with $E^2-B^2$ term will break time-reversal, and this may be the origin of the problem.} 
Presumably, in order to realize the impossible anomaly, more than the normal hydrodynamic degrees of freedom (i.e. Goldstone bosons for the Wess-Zumino action) would be necessary (see e.g. \cite{Lin:2011mr}\cite{Neiman:2011mj}\cite{Bhattacharyya:2012xi}\cite{Manes:2019fyw}).

One remaining puzzle is what happens if the charge lattice is not one-dimensional. If the charge lattice is not one-dimensional, we have magnetic as well as electric conserved current. At the same time, we do not expect the Bianchi identity (although as long as the associated current is not anomalous, we may still impose the Bianchi identity on the external source).  Without the Bianchi identity, we have one less equation to rewrite the divergence of the entropy current. This makes it  more difficult, probably impossible to assure the positivity of the entropy production. A resolution of this apparent difficulty is not clear at this moment but it could indicate the failure of locality \cite{PV}.






\section*{Acknowledgements}

This work is in part supported by JSPS KAKENHI Grant Number 17K14301. The author thanks J.~Bhattacharya for the correspondence and useful comments in this difficult situation all over the world.

\end{document}